\documentclass[twocolumn,11pt]{article}
\usepackage{lscape}
\usepackage{epsfig}
\usepackage{amssymb}
\usepackage{txfonts}
\usepackage{graphicx}
\usepackage{natbib}

\begin{document}

   \title{ Interstellar H${\bf_2}$ toward HD 37903 }  % \bmath
   %\title{ INTERSTELLAR H${\bf_2}$ TOWARD HD 37903}

   \author{ {\bf Gnaci\'nski Piotr } \\ % }{
   %\institute{
   %\\
   %\author{ P. Gnaci\'nski }
   %\author[P. Gnaci\'nski]{ P. Gnaci\'nski  $^1$ \thanks{email: pg@iftia9.univ.gda.pl} \\  
   % $^1$   
   %\email{pg@iftia9.univ.gda.pl}
   %\affil{
   Institute of Theoretical Physics and Astrophysics, 
              University of Gda\'nsk, \\
              ul. Wita Stwosza 57, 80-952 Gda\'nsk \\
              %\email {\it pg@iftia9.univ.gda.pl} 
               email: {\it pg@iftia9.univ.gda.pl}              
   }

   \date{\today}

   \maketitle
   
   \abstract{
   %\begin{abstract}
      We present an analysis of interstellar H$_2$ toward HD 37903, which is
      a hot, B 1.5 V star located in the NGC 2023 reflection nebula. %\cite{Meyer} 
      Meyer {\it et al.} (2010)
      have used a rich spectrum of vibrationally excited H$_2$ observed by the HST to calculate a model of the interstellar cloud toward HD 37903.
      We extend Mayer's analysis by including the $\nu$''=0 vibrational level observed by the FUSE satellite.
    
      The T$_{01}$ temperature should not be interpreted as a rotational temperature, but rather as a temperature of thermal equilibrium between the ortho and para H$_2$. The ortho to para H$_2$ ratio is lower for the lowest rotational levels than for the higher levels populated by fluorescence. 
      The PDR model of the cloud located in front of HD 37903 points to a gas temperature T$_{kin}$=110--377 K, hydrogen density n$_H$=1874--544 cm$^{-3}$ and the star--cloud distance of 0.45 pc.  
   %
   % model b900d45r1 izobaryczny z bilansem energetycznym
   %
   %\end{abstract}
   }
   
   %\keywords {
   {\bf  Key words: }{\it
   %\begin{keywords}
    ISM: clouds --- ISM: molecules --- ultraviolet: ISM 
   %\end{keywords}
   }
   %\end{abstract}
   
   %\maketitle

\section{Introduction}
%\section{INTRODUCTION}

 A rich spectrum of vibrationally excited H$_2$ in the direction to HD 37903 was first described by 
\cite{Meyer}. They have observed over 500 interstellar H$_2$ absorption lines from
excited vibrational levels $\nu$"=1--14 and rotational levels up to J"=13. These lines were detected in
a {\it Hubble Space Telescope} (HST) spectrum made with the {\it Space Telescope Imaging Spectrograph} (STIS).
 A {\it Far Ultraviolet Spectroscopic Explorer} (FUSE) spectrum was made after Mayer's publication 
allowing to access the $\nu$"=0 vibrational level of the ground electronic state. The FUSE spectrum
was used by \cite{Rachford} to determine the T$_{OP}=170\,K/\ln({9N_0}/{N_1})$=68 $\pm$ 7 K gas "kinetic" temperature
and the hydrogen molecular fraction f(H$_2$)=2N(H$_2$)/(2N(H$_2$)+N(HI))=0.53 $\pm$ 0.09 in the direction towards HD 37903.

 \begin{figure*}%[tb]
    \centering
    \includegraphics[width=\textwidth,viewport=1 1 840 500,clip]{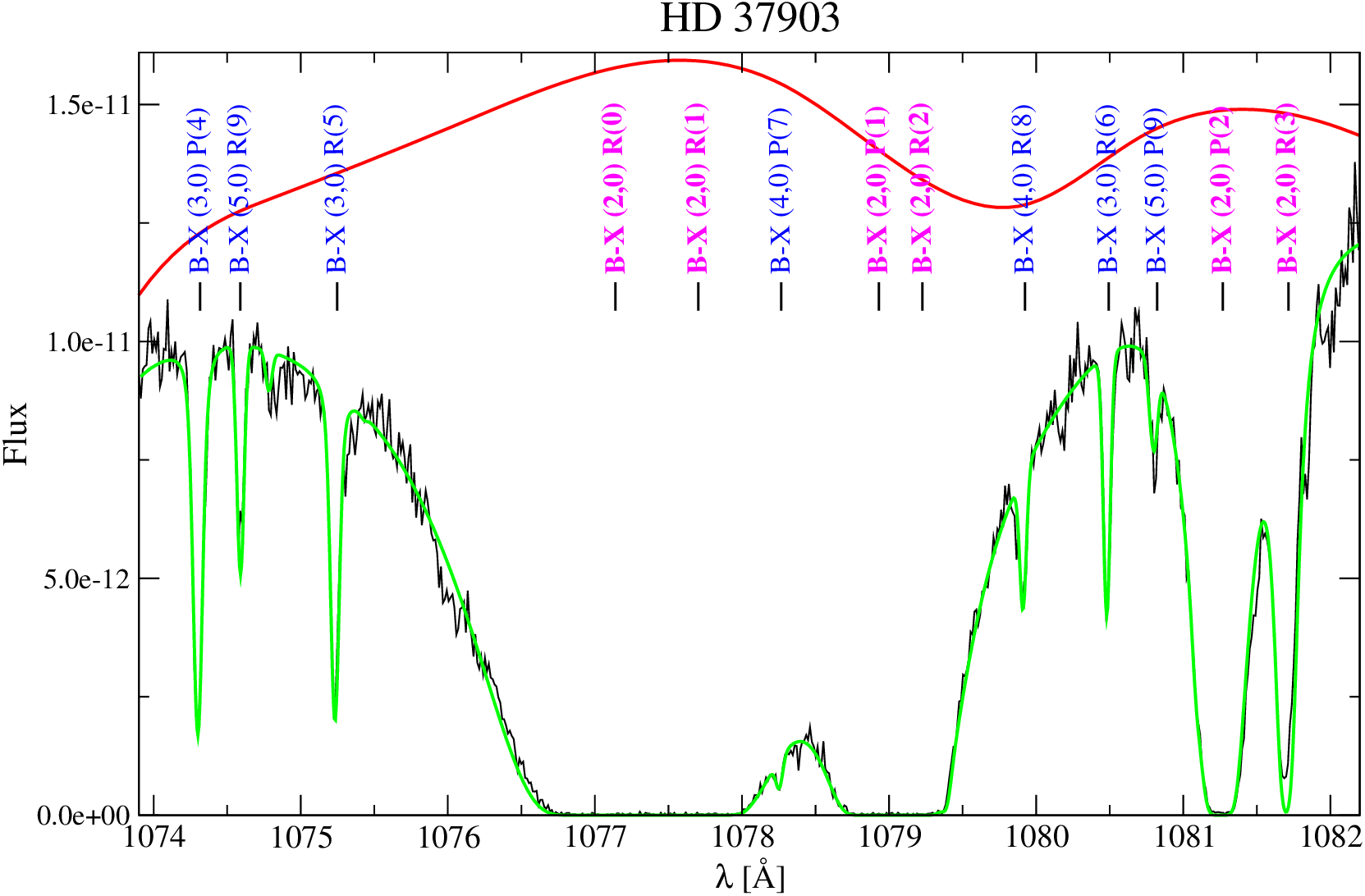}
    \caption{ A fragment of FUSE spectrum with a fit (green line) of H$_2$ absorption lines.
      The red line represents continuum.    }
    \label{fuse}
 \end{figure*}

The star HD 37903 was also observed by the {\it Berkeley Extreme and Far--Ultraviolet Spectrometer} (BEFS) onboard
the {\it Orbiting and Retrievable Far and Extreme Ultraviolet Spectrometer} (ORFEUS) telescope.
The spectral resolution was R=3000. \cite{Lee} have used this spectra to obtain  H$_2$ column
densities on $\nu$"=0 and J"=0--5 rotational levels. Their column densities agree within the order of magnitude to the column densities derived in this paper from the FUSE spectra. The physical parameters derived by \cite{Lee} are: T$_{OP}$=63 $\pm$ 5 K, f(H$_2$)=0.496 $\pm$ 0.017 and the cloud density
n=5600 cm$^{-3}$.

The H$_2$ molecule exists in two forms: 
ortho (odd J") and para (even J") H$_2$. It is caused by the spins of the hydrogen nuclei which can
point "in the same" direction (ortho H$_2$ -- triplet state,  total nuclei spin I=1) or in opposite directions (para H$_2$ -- singlet state, I=0).
The ratio ortho/para H$_2$ is therefore 3:1 at standard temperature and pressure.
Conversion between this two spin isomers can take place in gas phase, or on the surface of
gas grains \citep{Bourlot}. The ortho--para conversion in the gas phase is caused by the exchange of
proton in collisions with H, H$^+$ and H$_3^+$.

  The fluorescence cascade of H$_2$ that leads to population of excited ro-vibrational states of H$_2$ 
as well as to the emission of infrared photons from quadrupole transitions has been described
by \cite{Black}.
The first ultraviolet detection of vibrationally excited interstellar H$_2$ was performed by 
\cite{Federman} in the HST spectrum of $\zeta$ Ophiuchi.

 \begin{figure*}%[tb]
    \centering
    \includegraphics[width=\textwidth,viewport=1 1 832 470,clip]{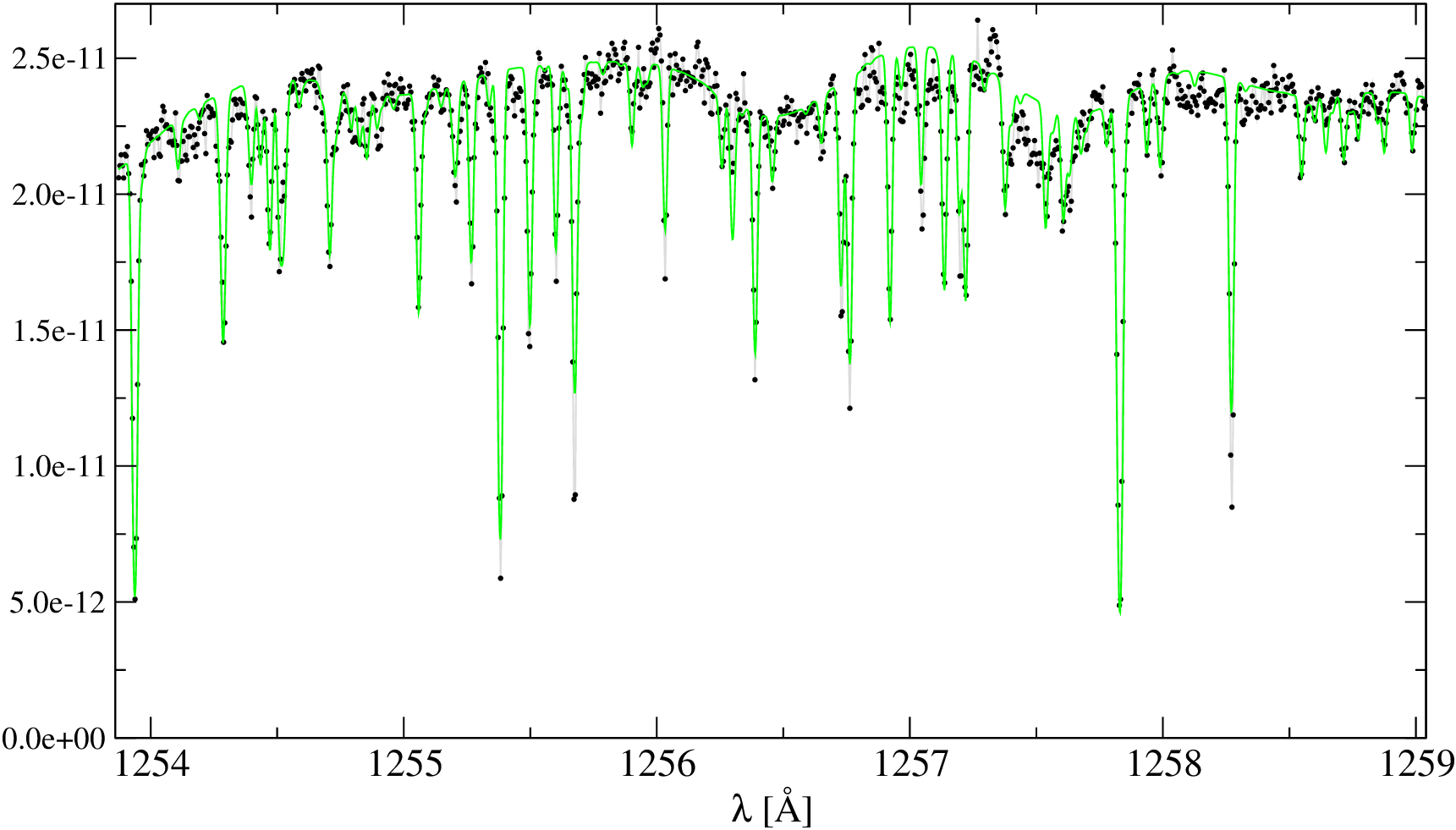}
    \caption{ Fragment of HST STIS spectrum (gray line with dots) fitted with 268 H$_2$ absorption lines (green line). The figure presents only a 5 \AA\ fragment of the spectrum, while the presented fit was done to the whole 200 \AA\ long STIS spectrum (1160--1357 \AA) and included 7449 H$_2$ lines.  }
    \label{HSTspectrum}
 \end{figure*}

\section{Column densities}
%\section{COLUMN DENSITIES}

 \begin{figure*}%[tb]
    \centering
    \includegraphics[width=\textwidth,viewport=1 1 840 530,clip]{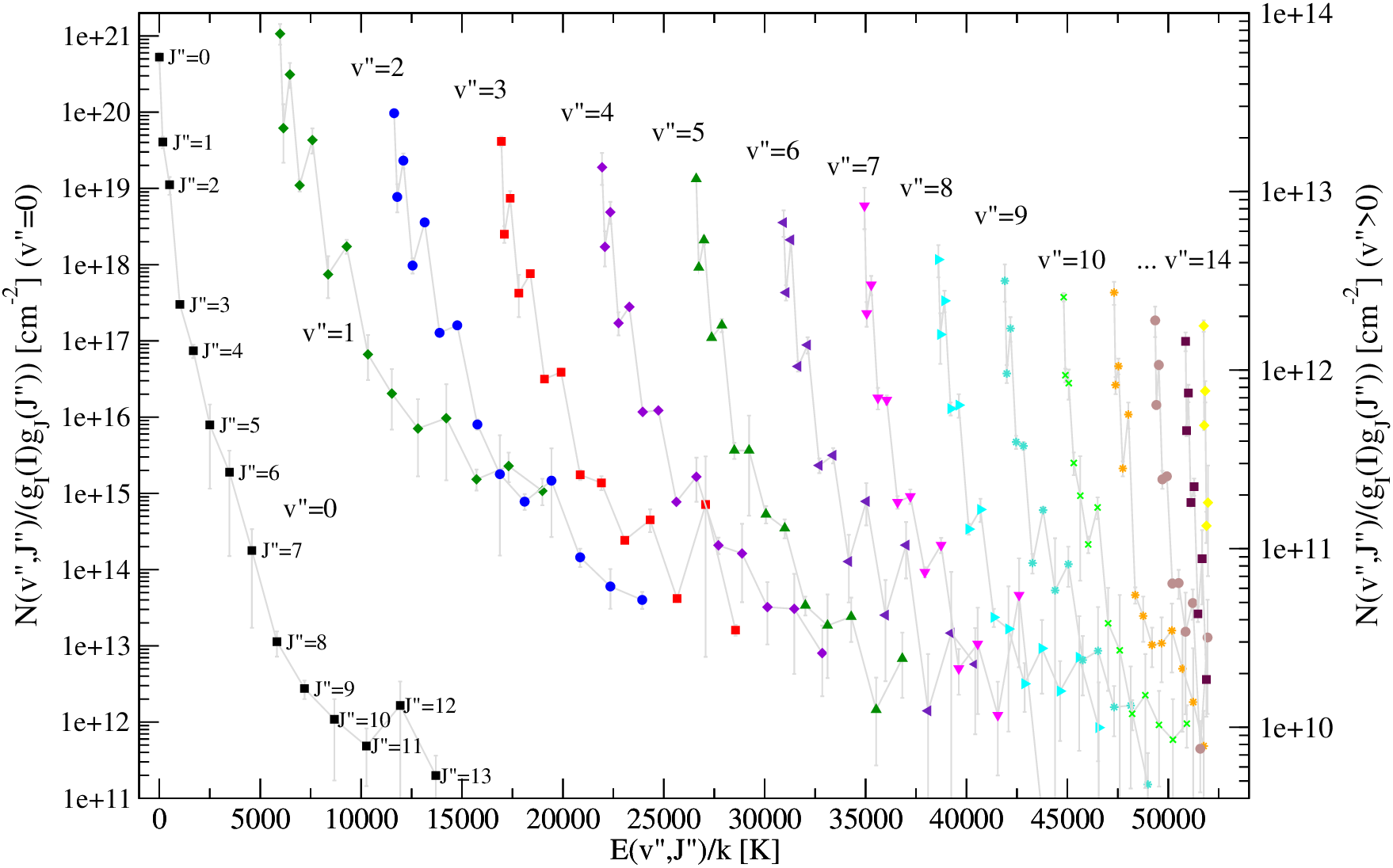}
    \caption{ Occupation of H$_2$ X ro-vibrational levels towards HD 37903.  
             Note the different y-axis scale for $\nu$"=0 (left) and for higher $\nu$" levels (right). 
             %The lines show the least square linear fit to ortho and para H$_2$ rotational levels used to calculate the rotational temperature of the $\nu"=2$ vibrational level .  
           }
    \label{H2levels}
 \end{figure*}
 
  We have used both HST STIS and FUSE spectrum to obtain column densities on H$_2$ ro-vibrational levels. 
  The HST STIS spectrum o59s04010 was averaged because it consist of two subexposures.
  The HST spectrum is located between 1160 and 1357 \AA. 
We have also used the FUSE observation P1160601.
We have analyzed only spectrum from detectors 1B LiF and 1A LiF. These two spectra had
the best quality. 
The FUSE spectra originating from the same detector were shifted and coadded using the IRAF tasks 
{\it poffsets} and {\it specalign}. 
A part of the FUSE spectrum is shown on fig. \ref{fuse}.
The whole FUSE spectrum ranges from 987 \AA\ to 1188 \AA.

The Ar I 1048 \AA\ line that lies in the wing of R(0) and R(1) lines was cut out from the spectrum. 
The wings of H$_2$ lines were calculated up to 60 \AA\ from the line center. The absorption lines in the
FUSE spectrum were modeled with the Voigt function, while a Gauss function was used for modeling
the H$_2$ lines in the STIS spectrum.
  
We have used a Gaussian point spread function (PSF) with  FWHM equal to 15 km/s \citep{Jensen} for modeling the FUSE spectra.
%$\sigma=0.03$ \AA\ for the FUSE spectrum.
%The spectral resolution R=20000 given by \cite{Andersson} leads to $\sigma=0.05$ \AA\ for the Gaussian PSF at 1000 \AA. However, during the fitting procedure we have found that this value is to large.
The column densities at J"=0, 1 and 2 ($\nu$"=0) were derived from transitions between B and X electronic levels. The vibrational transitions used are
(0,0), (2,0), (3,0) and (4,0), where the first digit is the $\nu'$ on the upper electronic level B and the second digit $\nu"=0$ is the vibrational level of the ground electronic state X. 
The FUSE spectrum at the (1,0) vibrational transition was to noisy to
perform a good fit.
 The H$_2$ line positions and oscillator strengths were adopted from \cite{Abgrall}.
The total transition probabilities that include the transitions to continuum (dissociating) states were taken from \cite{Abgrall-Atotal}.

% Sprawdzic czy nie uzylem linii H2 1025 \AA\ (B-X (6,0) R(0)) zblendowanej z Lyman beta !!! - NIE !!!

\begin{table*}
\centering
\caption{Column densities of the H$_2$ ro-vibrational levels towards HD 37903 [cm$^{-2}$].}
\label{CD}
  \tiny
  \scriptsize
\begin{tabular}{rrrrrrrrrrrrrrrr}
\hline
J"$\backslash\nu$" & 0 & 1 & 2 & 3 & 4 & 5 & 6 & 7 & 8 & 9 & 10 & 11 & 12 & 13 & 14 \\
\hline
0 & 5.3e20 & 7.7e13 & 2.7e13 & 1.9e13 & 1.4e13 & 1.2e13 & 6.7e12 & 8.3e12 & 4.2e12 & 3.2e12 & 2.6e12 & 2.7e12 & 1.9e12 & 1.5e12 & 1.8e12 \\
1 & 3.7e20 & 2.0e14 & 8.4e13 & 5.2e13 & 4.4e13 & 3.4e13 & 2.4e13 & 1.9e13 & 1.4e13 & 8.6e12 & 8.4e12 & 7.4e12 & 5.7e12 & 4.1e12 & 4.4e12 \\
2 & 5.6e19 & 2.3e14 & 7.4e13 & 4.6e13 & 3.8e13 & 2.7e13 & 2.7e13 & 1.5e13 & 1.2e13 & 8.6e12 & 4.2e12 & 5.3e12 & 5.3e12 & 3.7e12 & 3.8e12 \\
3 & 6.3e18 & 2.3e14 & 8.1e13 & 5.7e13 & 3.9e13 & 3.2e13 & 2.2e13 & 1.5e13 & 1.3e13 & 8.3e12 & 6.3e12 & 5.9e12 & 5.1e12 & 3.8e12 & 2.8e12 \\
4 & 6.7e17 & 1.8e14 & 6.1e13 & 3.1e13 & 2.0e13 & 1.6e13 & 1.2e13 & 6.1e12 & 5.7e12 & 3.4e12 & 1.8e12 & 5.1e12 & 2.3e12 & 2.0e12 & 1.6e12 \\
5 & 2.6e17 & 1.1e14 & 5.3e13 & 2.9e13 & 1.9e13 & 1.2e13 & 9.7e12 & 6.0e12 & 4.2e12 & 2.7e12 & 3.5e12 & 1.8e12 & 2.1e12 & 1.4e12 & --- \\
6 & 2.5e16 & 6.4e13 & 2.3e13 & 1.3e13 & 7.7e12 & 4.6e12 & 4.3e12 & 2.6e12 & 2.2e12 & 2.1e12 & 2.2e12 & 5.5e11 & 8.3e11 & 1.1e12 & --- \\
7 & 8.0e15 & 5.5e13 & 2.2e13 & 1.2e13 & 8.2e12 & 7.0e12 & 3.8e12 & 3.3e12 & 1.9e12 & 2.6e12 & 1.7e12 & 1.3e12 & 1.5e12 & 8.3e11 & --- \\
8 & 1.9e14 & 1.3e13 & 4.4e12 & 4.0e12 & 4.3e12 & 2.2e12 & 3.1e12 & 1.8e12 & 6.0e11 & 1.4e12 & 4.6e11 & 5.0e11 & 8.4e11 & --- & --- \\
9 & 1.6e14 & 2.7e13 & 1.0e13 & 6.4e12 & 5.9e12 & 2.7e12 & 2.4e12 & 1.2e12 & 1.0e12 & 1.4e12 & 6.8e11 & 2.0e12 & 4.3e11 & --- & --- \\
10 & 2.3e13 & 1.1e13 & 5.0e12 & 3.0e12 & 2.0e12 & 7.8e11 & 2.2e12 & 6.2e11 & 5.8e11 & 5.6e11 & 3.2e11 & 4.5e11 & 6.7e11 & --- & --- \\
11 & 3.3e13 & 1.7e13 & 6.2e12 & 3.6e12 & 3.2e12 & 2.9e12 & 8.5e11 & 8.1e11 & 1.1e12 & 8.9e11 & 7.1e11 & 9.5e11 & --- & --- & --- \\
12 & 4.1e13 & 7.3e12 & 1.5e12 & 4.4e12 & 1.2e12 & 3.1e11 & 8.4e11 & 1.4e12 & 6.2e11 & 3.3e11 & 2.1e11 & 2.0e11 & --- & --- & --- \\
13 & 1.6e13 & 1.7e13 & 4.2e12 & 2.8e12 & 2.1e12 & 2.0e12 & 1.8e12 & 3.0e11 & 8.1e11 & 3.9e11 & 8.5e11 & --- & --- & --- & --- \\
\hline
\end{tabular}
\end{table*} 
 
  The column densities of rotational levels J"=0--9 from the vibrational $\nu$"=0 level as well as 
the column density of the $\nu$"=1 J"=0 level were derived from the FUSE spectrum. 
Other ro-vibrational levels of the X ground electronic state were derived from the o59s04010 HST STIS spectrum.

  The column densities were derived with the profile fitting technique. At each point of the simulated spectrum the optical depth of many spectral lines has been summed. Such procedure was necessary to calculate the profile of blended lines which were ubiquitous in the spectra.
The cloud velocity, the doppler broadening parameter and column densities on all observed levels were free parameters, that were fitted to the observed spectra.

The STIS spectrum was fitted with H$_2$ absorption lines from all vibrational $\nu$"=0--14 levels, 
and rotational levels J"=0--13.
H$_2$ lines that were blended with atomic lines were excluded from the fitting procedure.
Total 7449 H$_2$ lines were included in the simulated STIS spectrum.
The whole STIS spectrum (200 \AA\ long) was fitted at once with all 7449 H$_2$ lines, because of large number of blended lines.
A fragment of the HST STIS spectrum with lines of vibrationally excited molecular hydrogen is presented on Fig. \ref{HSTspectrum}.
The observed column densities are presented in Table \ref{CD} and on Fig. \ref{H2levels}. The errors of the column density are about
20\% for J"=0--7 rotational levels, and up to $\sim$40\% for higher rotational levels.

\section{Results}
%\section{RESULTS}
  
 Total observed column density of H$_2$ on all vibrational and rotational levels equals 
to N(H$_2$)=($9.6\pm1.6)\cdot10^{20}$ cm$^{-2}$. 
The observed H$_2$ column density is slightly higher than N(H$_2$)=$4\cdot10^{20}$ cm$^{-2}$ predicted by \cite{Meyer}.
Column density of neutral hydrogen N(HI)=$1.48\cdot10^{21}$ cm$^{-2}$
\citep{Diplas}. The hydrogen molecular fraction (assuming 10\% error of N(HI)) equals to 
f(H$_2$)=2N(H$_2$)/(2N(H$_2$)+N(HI))=0.56$\pm$0.04.
% Additional doppler components are present in the O I 1356 \AA\ line or in the S II 1250 \AA\ one.
% They are not seen in the H$_2$ or C I spectral lines. The additional doppler components are probably from H II regions and do not influence the observed H I column density.

 \begin{figure}%[tb]
    \centering
    \includegraphics[width=\columnwidth,viewport=1 1 710 490,clip]{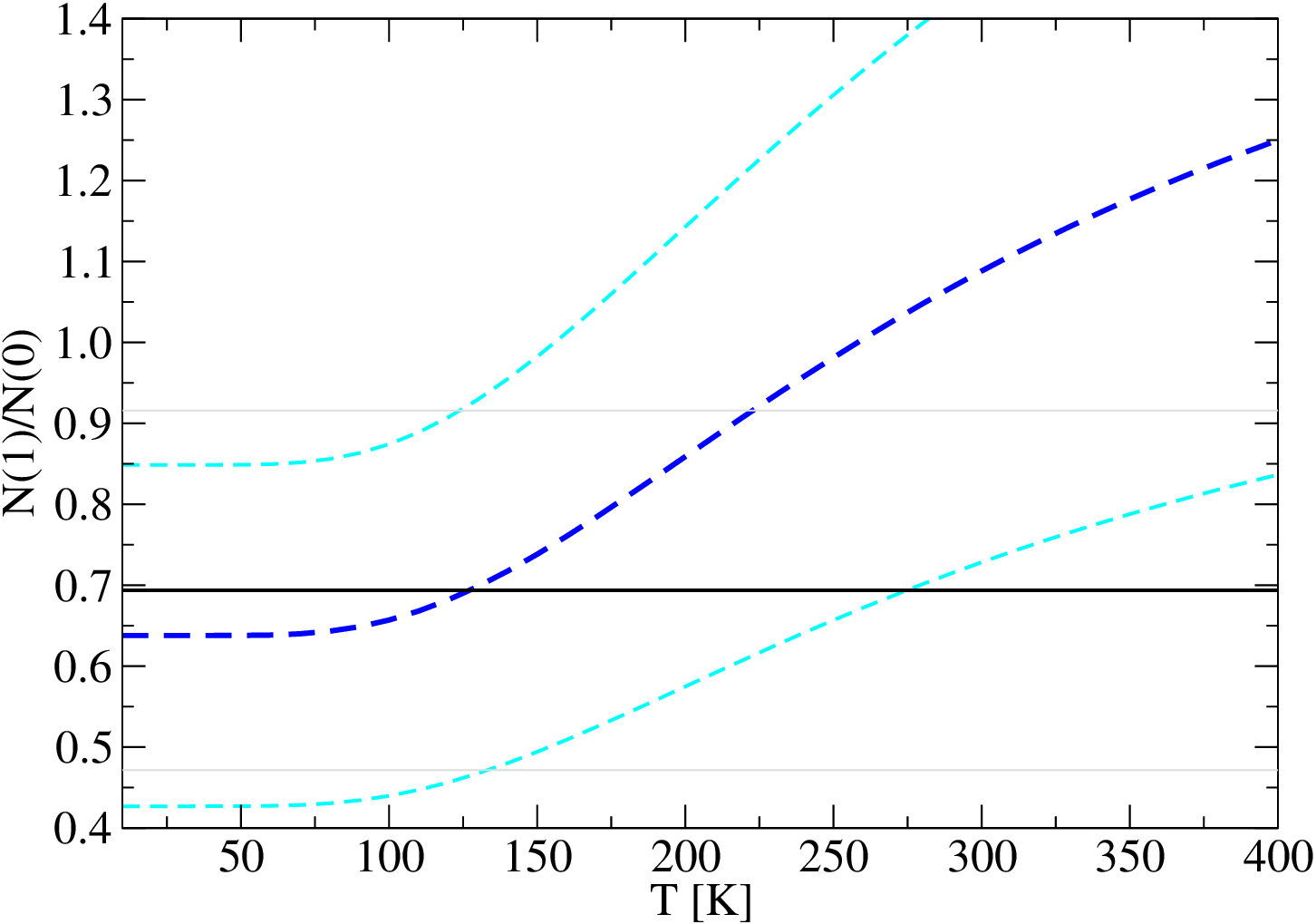}
    \caption{ Theoretical N(J=1)/N(J=0) ratio on the $\nu$"=0 vibrational level from eq. \ref{eq_OP} is shown as a dashed blue line. 
             The light blue dashed lines represent errors introduced by the uncertainty of the N$_{ortho}$/N$_{para}$ H$_2$ ratio 0.64 $\pm$ 0.21. 
             The observed N(1)/N(0) ratio together with its uncertainty is plotted as a straight line. }
    \label{fig_T01}
 \end{figure}

 \begin{figure}%[tb]
    \centering
    \includegraphics[width=\columnwidth,viewport=1 1 696 470,clip]{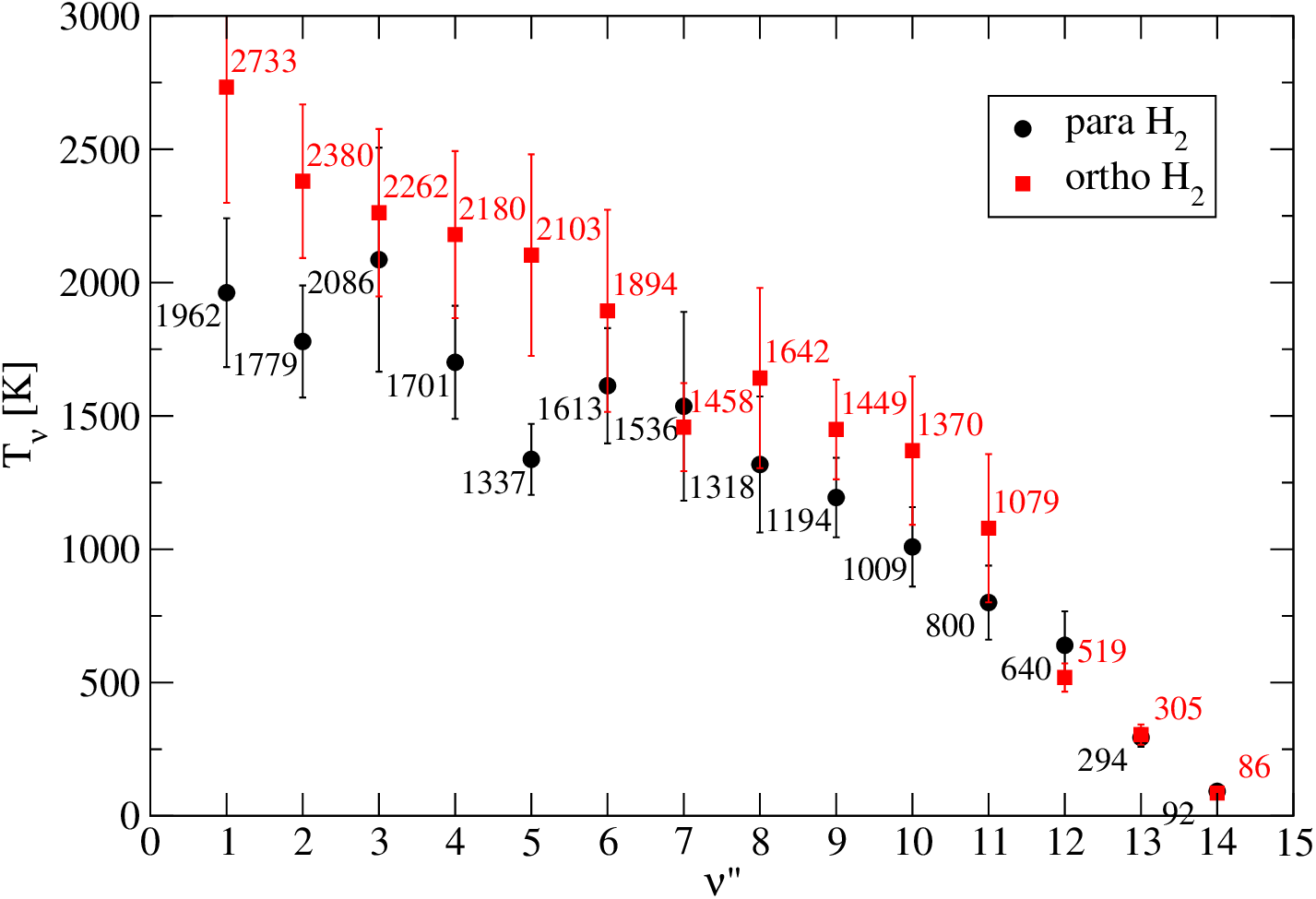}
    \caption{ Rotational temperatures for para and ortho H$_2$ toward HD 37903 as a function  of the vibrational number $\nu"$.}
    \label{fig_Trot}
 \end{figure}
 
\begin{table}
\centering
\caption{ Rotational temperatures for para and ortho H$_2$ toward HD 37903. }
\label{tab_Trot}
\begin{tabular}{rrrrrrr}
\hline
$\nu"$ & \multicolumn{2}{c}{T$_{para}$  [K]}   & \multicolumn{2}{c}{T$_{ortho}$ [K]}  \\
\hline
1	& 1962	&$\pm$	279	& 2733	&$\pm$	434 \\
2	& 1779	&$\pm$	210	& 2380	&$\pm$	288 \\
3	& 2086	&$\pm$ 420	& 2262	&$\pm$	314 \\
4	& 1701	&$\pm$ 212	& 2180	&$\pm$	313 \\
5	& 1337	&$\pm$	133	& 2103	&$\pm$	378 \\
6	& 1613	&$\pm$ 216	& 1894	&$\pm$	379 \\
7	& 1536	&$\pm$ 354	& 1458	&$\pm$	165 \\
8	& 1318	&$\pm$ 255	& 1642	&$\pm$	338 \\
9	& 1194	&$\pm$	149	& 1449	&$\pm$	187 \\
10	& 1009&$\pm$	149	& 1370	&$\pm$	278 \\
11	& 800	&$\pm$	139	& 1079	&$\pm$	278 \\
12	& 640	&$\pm$	127	& 519		&$\pm$	53 \\
13	& 294	&$\pm$	35	& 305		&$\pm$	38 \\
14	& 92	&$\pm$	3		& 86		&$\pm$	9 \\
\hline
\end{tabular}
\end{table} 

The T$_{OP}$ temperature, calculated from the lowest energy levels of ortho and para H$_2$ ($\nu"=0$) was derived from the equation:
\begin{equation} \label{eq_T01}
 \frac{N(1)}{N(0)}=\frac{ g_I(1) g_J(1)}{g_I(0) g_J(0) } \exp{\left(-\frac{E(1)-E(0)}{kT_{OP}}\right)},
 \end{equation}
 where $g_I$ is the spin degeneracy factor $g_I(I)=2I+1$. The $g_I(1)=3$ is the statistical weight for ortho H$_2$, and $g_I(0)=1$ for the para--H$_2$ spin izomer. The $g_J(J)=2J+1$ is the statistical weight for the rotational level J.
The T$_{OP}$ is the temperature of thermal equilibrium between the ortho and para spin izomers.
The derived T$_{OP}$=67$\pm$8 K is similar to the temperatures determined by \cite{Rachford} (68 K) and by \cite{Lee} (63 K).
The photon dominated regions (PDR) models of interstellar clouds shows that the T$_{OP}$ temperature is correlated with the gas kinetic temperature \citep{Petit}.

However, if we want to calculate the rotational temperature across the ortho--para divide we have to take into account the N$_{ortho}$/N$_{para}$ H$_2$ ratio. 
The population of the ro-vibrational levels of H$_2$ depends not only from the temperature and
radiation field, but also from the total ortho to para ratio. In order to include the ortho to para ratio into
the Boltzmann equation we will write separate Boltzmann distributions for the ortho and para spin izomers.
The Boltzmann distribution for the ortho H$_2$ can be written as:
 \begin{equation} \label{eq_No}
 N_{o}(J_{o}")=\frac{N_{ortho}}{Z_o(T)} g_I(1) g_J(J_o") \exp{\left( -\frac{E(J_{o}")}{kT} \right)}
 \end{equation}
 where N$_{ortho}$ is the total amount of ortho H$_2$. 
 The partition function $Z$ is:
 \begin{equation}
 Z_o(T)=\sum_{J_o" \; odd} g_I(1) g_J(J_o") \exp{\left(-\frac{E(J_o")}{kT}\right)}.
 \end{equation}
 
 Let us assume, that para H$_2$ has a Boltzmann distribution with the same rotational temperature: 
 % Similar equation can be written for the para H$_2$:
 \begin{equation} \label{eq_Np}
 N_{p}(J_{p}")=\frac{N_{para}}{Z_p(T)} g_I(0) g_J(J_p") \exp{\left(-\frac{E(J_{p}")}{kT}\right)}
 \end{equation}
 where $N_{para}$ is the total amount of para H$_2$, and the partition function for para H$_2$:
 \begin{equation}
 Z_{p}(T)=\sum_{J_p" \; even} g_I(0) g_J(J_p") \exp{\left(-\frac{E(J_{p}")}{kT}\right)}.
 \end{equation}
 
 By dividing the equations \ref{eq_No} by \ref{eq_Np} we obtain:
 \begin{eqnarray}   
   \label{eq_OP}
 \frac{N_o(J_o")}{N_{p}(J_{p}")}=\frac{N_{ortho}}{N_{para}} \frac{ g_I(1) g_J(J_o")}{ g_I(0) g_J(J_p")}\frac{Z_p(T)}{Z_o(T)} \nonumber  \\  \cdot
  \exp{\left(-\frac{E(J_{o}")-E(J_{p}")}{kT}\right)}.
 \end{eqnarray}
We have tried to solve the above equation numerically for the J$_o"$=1 and J$_p"$=0 states ($\nu$"=0) to obtain the T$_{01}$ rotational temperature. 
 The observed $N_{ortho}/N_{para}=0.64\pm 0.21$.
 Figure \ref{fig_T01} presents the right side of eq. \ref{eq_OP} and the ratio of observed column densities (left side of eq. \ref{eq_OP}) together with maximal errors. 
 The rotational temperature T$_{01}$ can take values between 0 and 500 K.

  The equation \ref{eq_OP} can be written in another way by making two assumptions that are 
{\bf not valid} % THIS SHOULD BE LEFT BOLD !
in the interstellar medium:
\begin{enumerate}
	\item $T>240$ K $\Rightarrow Z_p(T)/Z_o(T) \approx 1/3$
	\item $N_{ortho}/N_{para}=3$ as in standard laboratory conditions.
\end{enumerate}
 Under this assumptions the eq. \ref{eq_OP} for rotational temperature reduces to:
  \begin{equation} 
  \label{tradycyjne}
 \frac{N_o(J_o")}{N_{p}(J_{p}")}=\frac{ g_I(1) g_J(J_o")}{ g_I(0) g_J(J_p")} \exp{\left(-\frac{E(J_{o}")-E(J_{p}")}{kT}\right)}.
 \end{equation}
 and $T_{01}$ is the same as the temperature $T_{OP}$ of ortho--para thermal equilibrium.
 %The difference between the rotational temperature and the ortho--para temperature can be significant in application of T$_{OP}$ as reference to DIB broadening because of unresolved rotational levels, like \cite{Maja}.
 
The rotational temperatures T$_{02}$ and T$_{13}$ involve rotational levels of the same spin isomers of H$_2$, and do not depend on the total N$_{ortho}$/N$_{para}$ H$_2$ ratio and on the partition function. 
For the cloud towards HD 37903 the T$_{02}=132\pm14$ K and T$_{13}=172\pm9$ K. 
However, the T$_{12}$ temperature calculated from eq. \ref{eq_OP} is  $133^{+37}_{-25}$ K, while the use of eq. \ref{tradycyjne} gives a incorrect temperature of $263 \pm 89$ K. 
 
  A good illustration that the eq. \ref{eq_OP} must be used can be the $\nu$=1 level. On the $\nu=1$ level the temperature T$_{12}=1223$ K (levels ortho and para) is close to the T$_{02}=921$ K and T$_{13}=1090$ K  temperatures calculated from the same spin isomers. The temperature calculated with the traditionally used equation \ref{tradycyjne} is negative T$=-466$ K ! 
 
  The ortho (odd J") to para (even J") H$_2$ ratio (hereafter O/P) $(O/P)_{\nu"=1-14}=1.35\pm0.18$ was calculated by summing ortho and para H$_2$ on vibrational levels $\nu$"=1--14. It is a bit lower than O/P=$1.45\pm0.08$ given by \cite{Meyer}. 
We have also used the method proposed by \cite{Wilgenbus} to calculate O/P H$_2$ from individual ortho levels,
specially  $J_o"$=1 and 3 on $\nu"$=0.
%The Wilgenbus {\it et al.} method allows to calculate the O/P H$_2$ ratio from a ortho level ($J_o")$ and two adjacent para H$_2$ levels: $J_o"-1$ and $J_o"+1$. First one should calculate the rotational temperature from the two para H$_2$ levels: 
%\begin{equation} 
%T_{rot}=\frac{E(J_o"+1)-E(J_o"-1)}{k \ln\frac{N(J_o"-1)g_J(J_o"+1)}{g_J(J_o"-1)N(J_o"+1)}}.
%\end{equation}
%Next we can calculate the O/P ratio: 
%\begin{equation} 
%  \label{Wilgenbus}
% \left(\frac{O}{P}\right)_{J_o"}=\frac{ Z_o(T_{rot})}{ Z_p(T_{rot})}\frac{N(J_o")}{N^{interp}(J_o")},
% \end{equation}
%where $N^{interp}(J_o")$ is the column density of an ortho state interpolated from the two adjacent para states on a $\ln(N(J)/(g_I(I)g_J(J)))$ against $E(J)/k$ diagram. 
For the $\nu$"=0 J"=1 ortho state the O/P ratio differs significantly from the $(O/P)_{\nu"=1-14}=1.35$ value and is equal to $(O/P)_{J"=1}=0.63\pm0.11$. Also for the $\nu$"=0 J"=3 ortho state the observed O/P is low - only $(O/P)_{J"=3}=0.70\pm0.07$. For higher rotational and vibrational levels the \cite{Wilgenbus} method gives (O/P)$_{J_o}$ ratio close to 1.3 derived for excited vibrational H$_2$ states.
 
 The rotational temperatures for vibrational levels $\nu"$=1--14 for para H$_2$ were obtained from the linearized Boltzmann distribution:
 \begin{equation}  \label{eq_Trot}
   \ln \frac{N_{p}(J_{p}")}{g_I(0)g_J(J_p")} =\ln{N_{p}(0)}-\frac{E(J_{p}")}{kT_\nu}.
 \end{equation}
Analogous equation was used for the distribution of the ortho H$_2$ states. 
Fig. \ref{H2levels} shows left side of eq. \ref{eq_Trot} versus E/k. 
Levels fulfilling the Boltzmann distribution should be placed on a straight line on this plot.
%The lines on fig.  \ref{H2levels} show the least square fit with eq. \ref{eq_Trot} to all ortho H$_2$ levels (J"=1,3,5,7,9,11,13) and separately (second line) to para H$_2$ rotational levels (J"=0,2,4,6,8,10,12) of the $\nu"=2$ vibrational level. 
On the $\nu$"=0 vibrational level, which is populated partially by collisions the occupation of
rotational levels does not follow a straight line. Therefore the rotational temperature was not calculated for the $\nu$"=0 vibrational level.
For both spin isomers and for vibrational levels $\nu$"=1--14 the temperature was calculated using the the linear regression method. 
The inverse of the line inclination ($-1/T_\nu$ in eq. \ref{eq_Trot}) gives us the rotational temperature. 
% taken with minus sight is the rotational temperature.
The resulting temperatures are presented 
in table \ref{tab_Trot} and 
on figure \ref{fig_Trot}. 
The rotational temperatures for the ortho isomer are usually higher than for the para H$_2$. 
However the ortho H$_2$ rotational temperatures calculated form levels J$_o$=1 to last but one J$_o$ ortho state are in conformity with the para H$_2$ rotational temperatures. 
So the assumption that the rotational temperatures $T_\nu$ are the same for ortho and para H$_2$ seems to be justified. 
The rotational temperatures falls linearly with the increasing vibrational level.
 
\section{Model}

  We have used the Meudon PDR code \citep{Petit} to calculate a model of the 
interstellar medium in the direction of HD 37903. 
The interstellar reddening E(B-V)=0$^m\!\!.$35 and R$_V$=5.5 was chosen to match the values used by \cite{Meyer}.
The radiation source is the HD 37903 star (B 1.5 V spectral type).
The model includes 300 H$_2$ ro-vibrational levels and uses exact computation of radiative transfer
in H$_2$ spectral lines. The model is isobaric with thermal balance.

\begin{table}
\centering
\footnotesize
\caption{Comparison of cloud models toward HD 37903.}
\label{models}
\begin{tabular}{lccc}
\hline
& this paper       & \citet{Meyer} & \citet{Lee} \\
\hline
telescope & HST STIS        & HST STIS     & BEFS ORFEUS \\ 
          & and FUSE        &     & \\
H$_2$ levels      & v"= 0 -- 14  & v"= 1 -- 14  & v"= 0 \\
                  & J"= 0 -- 13  & J"= 0 -- 13  & J"= 0 -- 5 \\
R$_V$             & 5.5          & 5.5          & 4.1 \\
n$_H$ [cm$^{-3}$] & 544 --1874   & 130 -- 8 800 & 5 600 \\
d [pc]            & 0.45         & 0.5          & 0.2 \\
T$_{kin}$ [K]             & 110 -- 377   & 400          & --- \\
\hline
\end{tabular}
\end{table} 

All our models were two-side models with interstellar radiation field on the observer side equal to
one `Draine' unit. The interstellar radiation field on the star side has been varied from 1 to 10$^4$ `Draine' units.
The best model was chosen among 6917 different models by minimising the sum:
\begin{equation}
  \sum_{v",J"} w_{v",J"}\left(\log \frac{N_{obs}(v", J")}{N_{obs}(0, 0)}-\log \frac{N_{model}(v",J")}{N_{model}(0,0)}\right)^2
\end{equation}
where the weights $w_{v",J"}$ were chosen so, that the levels populated by collisions J"=0--6 (v"=0) have the same influence on the final sum as the rest of the levels (v"$>$0 and v"=0 J"=7--13), that are populated by fluorescence.

  We have varied the star--cloud distance, hydrogen density and the interstellar radiation field on the star side in order to find a model that matches the observations.

Table \ref{models} presents of comparison of our best
model and models presented by \cite{Meyer} and \cite{Lee}. The star -- cloud distance of 0.45 pc,
that is responsible for the filling the fluorescence H$_2$ levels is similar in our model 
and in the model presented by \cite{Meyer} (d=0.5 pc). Our cloud kinetic temperature (connected
with the collisional--filled levels) equals to T$_{kin}$=377 K on the star side of the cloud and T$_{kin}$=110 K on the observer's side. The hydrogen density obtained in our best model changes from n$_H$=544 cm$^{-3}$ on the star side to 1874 cm$^{-3}$ on the observer's side.
The interstellar radiation field on both sides of our best cloud model is one `Draine' unit.

%\section{Discussion}

% The right side of eq. \ref{eq_OP} for $J_o"=1$ and $J_p"=0$ has a limit :
% \begin{equation}   \label{eq_limit}
% \frac{O}{P} \cdot \frac{9 \exp{\left(-\frac{E(1)}{kT}\right)}}{Z_{o}(T)} \cdot \frac{Z_{p}(T)}{ %\exp{\left(-\frac{E(0)}{kT}\right)}} \stackrel{T\rightarrow 0}{ \longrightarrow} \frac{O}{P}
% \end{equation}

 \section{Conclusions}
%\section{CONCLUSIONS}
  
  Column density of molecular hydrogen toward HD 37903 on all observed H$_2$ levels N(H$_2$)=$9.6\cdot10^{20}$ cm$^{-2}$. The hydrogen molecular fraction f(H$_2$)=0.56. The ortho/para H$_2$ ratio equals to 1.35 for the excited levels, but for the J"=1 ortho state ($\nu$"=0) the ortho to para ratio is only (O/P)$_{J=1}$=0.63.
 
  %The T$_{01}$ temperature should not be called a rotational temperature. 
  The T$_{OP}$ is a temperature of thermal equilibrium between the ortho and para spin isomers and is equal to 67 K. The rotational temperatures T$_{02}$=132 K and T$_{13}$=172 K.
The formula for rotational temperatures calculated across the ortho--para divide (like T$_{12}$) should include the (O/P)$_\nu$ H$_2$ ratio. 
   
  The best PDR model for the cloud toward HD 37903 gives a gas kinetic temperature T$_{kin}$=110--377 K, 
hydrogen density n$_H$=544--1874 cm$^{-3}$ and star -- cloud distance of 0.45 pc.

\section*{Acknowledgments}

%\begin{acknowledgements}
%\acknowledgements
  I would like to thank Herve Abgrall for providing the natural line widths for the H$_2$ lines.
  The research was supported by University of Gda\'nsk grant BW/5400-5-0336-0.
%\end{acknowledgements}

\bibliographystyle{aa}

\end{document}